\documentclass[doublecol]{epl2} 
\usepackage{amsmath, amstext, amsfonts, mathrsfs, psfrag, yfonts, amssymb, bbm} % Mathebibliotheken
\usepackage{dcolumn}
\usepackage{graphicx}   % need for figures
\usepackage{verbatim}   % useful for program listings
\usepackage{color}      % use if color is used in text
\usepackage{subfigure}  % use for side-by-side figures

\title{Interaction of atomic quantum gases with a single \\ carbon nanotube}

\author{
M. Fink\inst{1}\thanks{E-mail: \email{finkm@ph.tum.de}} \and 
T.-O. M\"uller\inst{1} \and 
J. Eiglsperger\inst{2}\thanks{Present address: numares GROUP, Josef-Engert-Stra{\ss}e 9, \newline D-93053 Regensburg, Germany} \and 
J. Madro{\~n}ero\inst{1,3} 
}
\shortauthor{M. Fink \etal}

\institute{                    
  \inst{1} Physik Department, Technische Universit\"at M\"unchen, D-85747 Garching, Germany\\
  \inst{2} Institut f\"ur Theoretische Physik, Universit\"at Regensburg, D-93040 Regensburg, Germany\\
  \inst{3} Faculty of Physics, Universit\"at Duisburg-Essen, D-47045 Duisburg, Germany
}
\pacs{34.50.-s}{Scattering of atoms and molecules}
\pacs{78.67.Ch}{Nanotubes}
\pacs{67.85.-d}{Ultracold gases, trapped gases}

\abstract{
We study inelastic processes in the hybrid quantum system constituted by a carbon nanotube (CNT) in contact with an ultracold quantum gas, such as a cloud of thermal atoms or a Bose-Einstein condensate (BEC). We present a parameter-free ab-initio approach for the loss rate based on the underlying scattering process, considering the two-dimensional character of the system as well as the exact Casimir-Polder potential. The predicted loss rates are in perfect agreement with recent experimental results, obtained both for a thermal cloud of rubidium atoms and for a BEC. For the trap loss of a thermal cloud, we find that retardation effects become important and contribute significantly, which emphasises the crucial role of the exact interaction potential.
}

\begin{document}
\maketitle
% Introduction
\section{Introduction}
In recent years advanced techniques of cooling and controlling single atoms or ions, as well as clouds or condensates, have successfully been combined with sophisticated solid-state techniques in the production of mesoscopic structures on the nanometer scale. The emerging variety of hybrid quantum systems constitutes an interesting new field of physics~\cite{Wallquist.PhysScr.T137.014001,Sorensen.PRL.92.063601,Tian.PRL.93.266403,Hammerer.103.063005}. 
Nanotubes, nanowires and nanorods have proven effective for constructing such hybrid systems~\cite{Baughman.Science.297.787,Husain.ApplPhysLett.83.1240,Salem.NewJPhys.12.023039}. These can be combined into carpets of dense standing nanotubes forming a structured surface~\cite{Gierling.NatureNano.6.446,Pasquini.PRL.97.093201} or to an array of nanotubes serving as a diffraction grating~\cite{Juffmann.PRL.103.263601,Arndt.Nature.401.680}. Particular attention is currently being given to the fundamental system of a single nanotube interacting with a cold gas of atoms~\cite{Schneeweiss.NNano.2012.93,Fink.PRA.85.040702,Fink.PRA.81.062714,Jetter.arXiv.6161,Fermani.PRA.75.062905}, which lays the foundation for understanding systems of higher complexity.
This is by far not a trivial problem and raises the question of the applicability of established scattering theory and non-dynamical approaches~\cite{Jetter.arXiv.6161}.

\begin{figure}
\onefigure[width=75mm]{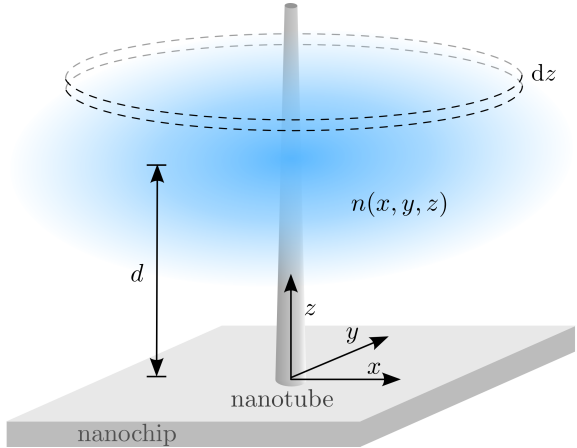}
\caption{(Color online) Schematic view of a CNT immersed in a cloud of trapped atoms. The nanotube is grown on top of a nanochip and the center of the cloud is held at a distance $d$ to this surface.}
\label{fig:schemetic view}
\end{figure}

In a recent experiment~\cite{Schneeweiss.NNano.2012.93,Schneeweiss.Diss} Schneeweiss {\em et al.} measured the losses of atoms absorbed by a multiwall carbon nanotube (length $L=10.25\mu$m and diameter ranging from $275$nm at the bottom to $40$nm at the tip) grown on top of a nanochip. The CNT was immersed into a BEC and in a cloud of thermal atoms and the trap loss $\gamma$ was measured as a function of the distance $d$ between the center of the trap and the surface of the nanochip (see fig.~\ref{fig:schemetic view}). 

In this Letter we present a parameter-free ab-initio description for the absorption of ultracold atoms by cylindrical geometries, based on the underlying scattering process. We show that an accurate description is possible, if the exact interaction potential is considered and the scattering process is treated properly. However, this remains non-trivial in the present case, due to the two-dimensional character of the scattering process~\cite{Lapidus.AJP.50.45,Verhaar.JPA.17.595,Adhikari.AJP.54.362} and due to the intricacy of the exact interaction potential~\cite{Fink.PRA.81.062714,Fink.EPJD.63.33}. In contrast to fitting an arbitrarily chosen model potential to the experimantal data~\cite{Schneeweiss.NNano.2012.93,Schneeweiss.Diss}, this parameter-free approach offers a foundation for extensions to more complex hybrid systems and might assist the design of CNT based nanodevices.\\

% Atom-Cylinder Potential
\section{The Casimir-Polder potential}
Descriptions of an atom interacting with a cylindrical geometry go back to Zel'dovich~\cite{Zeldovich.JETP.5.22} who analyzed the interaction between an atom and a perfectly conducting cylinder in 1935. Further work on this subject extended the potential to the the more realistic case of a cylinder with finite conductivity. A closed form of the full Casimir-Polder potential between an atom and a dielectric cylinder of radius $R$ was first given by Nabutovski \textit{et al.} in 1979~\cite{Nabutovski.JETP.50.352} and later by Marvin and Toigo in 1982~\cite{Marvin.PRA.25.782}. Though this problem has been revisited over the last years by several groups~\cite{Boustimi.PRB.67.045407,Bezerra.EPJC.71.1614,Eberlein.PRA.80.012504+PRA.75.032516}, the result obtained by Nabutovski \textit{et al.}~\cite{Nabutovski.JETP.50.352} is still the most general, as it covers the full Casimir-Polder potential of an atom interacting with a cylindrical geometry with arbitrary dielectric properties, including the perfectly 
conducting case.

The Casimir-Polder potential for a dielectric cylinder exhibits a nontrivial transition from the well-studied atom-wall potential at small atom-surface distances $(r-R)$, which is $-C_3/(r-R)^3$, to the fully retarded asymptotic $-C_6/r^6$ behavior far away from the cylinder. The coefficients $C_3$ and $C_6$, are well defined quantities depending on both the dielectric constant of the tube and the properties of the atom, with $C_6$ additionally depending on the radius $R$ of the cylinder~\cite{Nabutovski.JETP.50.352}. 
The full Casimir-Polder potential is approximated by the van der Waals potential only at very small distances and might resemble the van der Waals long-range asymptote $-C_5/r^5$ only in a very narrow transition region.
An accurate method for the numerical treatment of the full Casimir-Polder potential and, in particular for the nontrivial transition between the van der Waals and highly retarded limit, has only been developed very recently~\cite{Fink.PRA.85.040702,Fink.PRA.81.062714,Fink.EPJD.63.33}.\\

% Real Nanotube System
The interaction potential between an atom and an actual CNT of finite length $L$ obviously differs from the atom-cylinder potential~\cite{Nabutovski.JETP.50.352} in a nontrivial fashion and remains in general unknown; pairwise-summation approaches~\cite{Hamaker.Physica.4.1058} can in general not reproduce the correct Casimir-Polder forces~\cite{Rodriguez-Lopez.PRA.80.022519}. Variations of the interaction energy along the tube axis are, however, small as long as the radius $R(z)$ of the CNT  varies smoothly. In this case, the longitudinal free motion of an atom along the CNT may be separated from the two-dimensional dynamics perpendicular to the tube, which is governed by the interaction potential~\cite{Nabutovski.JETP.50.352} of the atom with a cylinder of radius $R(z)$.
The contribution of a single perpendicular plane to the total loss rate $\gamma$ can be given in a differential form
\begin{equation}
 \mathrm{d}\gamma(z) = n(0,0,z) K_\mathrm{in}^\mathrm{2D}\mathrm{d}z\,,
\end{equation}
with the density of particles $n$ around the nanotube, located at ($0$,$0$)  and the loss rate constant $K_\mathrm{in}^\mathrm{2D}$ for inelastic and reactive scattering in this two-dimensional (2D) subsystem.
These inelastic reactions (sticking, adsorption,~...) happen at short distances to the surface and are described in the Langevin model where all atoms reaching the surface contribute to the rate constant $K_\mathrm{in}^\mathrm{2D}$~\cite{Langevin.AnnCP.5.245,Julienne.Faraday.142.361}.
The total trap loss $\gamma$ of the full system is obtained by integration over all 2D planes,
\begin{equation}\label{eq:gamma total}
 \gamma = \int_{0}^L n(0,0,z) K_\mathrm{in}^\mathrm{2D}\mathrm{d}z\,.
\end{equation}
Deviations from this description might be expected at the tip of the nanotube where on the one hand, the dynamics along the tube axis is influenced by the atom-cylinder potential and on the other hand, the potenital differs from the infinite cylinder potential. However, these deviations will only give small corrections to the total trap loss.\\

% Introducing the Experiment
Within this description, we are able to calculate the trap loss of a Bose-Einstein condensate (BEC) and of a thermal cloud of atoms interacting with a CNT, as measured for both cases by Schneeweiss {\em et al.}~\cite{Schneeweiss.NNano.2012.93}. All calculations presented below are based on the exact experimental parameters~\cite{Schneeweiss.Diss,Gierling.Diss}\footnote{The trap frequencies and temperature of the cloud slightly change with the distance~\cite{Schneeweiss.Diss,Gierling.Diss}.}. The variation of the CNT radius $R(z)$ is assumed to be linear. The center of the cloud is shifted away from the tube in $y$-direction and lies on a ray forming an angle of $13^\circ$ with the $z$-axis~\cite{privcomm}. Furthermore, the experimental data is, for both cases (BEC and thermal cloud), assigned to trap-surface distances measured separately with a BEC\footnote{Notice that the authors of ref.~\cite{Schneeweiss.NNano.2012.93} relate their data for the thermal cloud to a different calibration~\cite{Schneeweiss.
Diss}, imposing a shift of $2.05\mu$m on the trap-surface distance.}.\\

% Bose-Einstein Condensate
\section{Absorption of a Bose-Einstein condensate}
\begin{figure}
\onefigure[width=75mm]{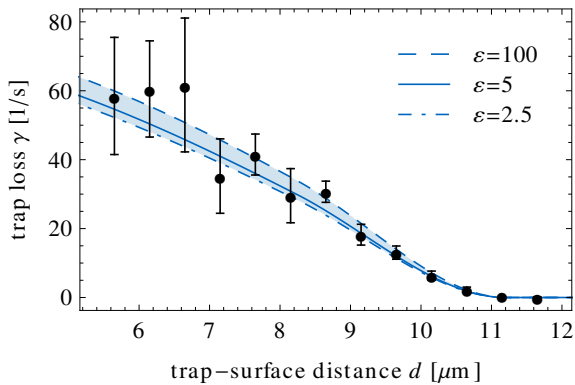}
\caption{(Color online) Trap loss $\gamma$ for a BEC of rubidium atoms overlapping with a carbon nanotube. The calculated trap-loss for a dielectric constant $\varepsilon=5$ (solid line) reproduces the data given in~\cite{Schneeweiss.NNano.2012.93} (full circles and errorbars). The dot-dashed (dashed) line shows results for a dielectric constant of $\varepsilon=2.5$ ($\varepsilon=100$) which leads to rather small variations in the predicted loss rates, compared to the experimental error bars.}
\label{fig:bec}
\end{figure}

Figure~\ref{fig:bec} shows the trap loss $\gamma$ for a BEC of rubidium atoms as a function of the trap-surface distance $d$. 
The BEC is described by a macroscopic wave function in the ground state of the trap and the presence of an absorbing impurity in the cloud leads to a local depletion of its density. As a result, the condensate readjusts with the characteristic speed of sound $v_s$ (here $0.8$mm$/$s)~\cite{Bloch.RevModPhys.80.885} which causes a flux toward the nanotube~\cite{Zipkes.nature.464.388}. Due to the symmetry in the region close to the nanotube this flux is symmetric around the tube axis. Only $s$-waves ($m=0$) reflect this symmetry and contribute to the absorption cross section $\sigma_\mathrm{abs}$. An additional factor $2\pi$ accounting for the isotropy of the incoming flux has to be considered and the loss rate constant is
\begin{equation}\label{eq:abs_cs bec}
 K_\mathrm{in}^\mathrm{2D} = 2\pi\,v_{s}\sigma_\mathrm{abs}^{(m=0)}(v_{s})\,.
\end{equation}
The density of the BEC in a harmonic trap is given by a Thomas-Fermi distribution~\cite{Bloch.RevModPhys.80.885}, with the characteristic Thomas-Fermi radii, $r^{\mathrm{TF}}_x=16\mu$m, $r^{\mathrm{TF}}_y=r^{\mathrm{TF}}_z=3.2\mu$m.
As the number of atoms in the condensate decreases during the absorption process, the Thomas-Fermi radii and the speed of sound do not stay constant over time, which leads to an algebraic instead of an exponential decay~\cite{Zipkes.nature.464.388}. These corrections and possible fluctuations due to collective oscillations~\cite{Schneeweiss.NNano.2012.93,Jetter.arXiv.6161} lie within the experimental error bars. The influence of the nanochip on the trap loss can be neglected for distances $d$ larger than $5\mu$m~\cite{Schneeweiss.NNano.2012.93}.
The cross section $\sigma_\mathrm{abs}^{(m=0)}$ is calculated using well established scattering theory~\cite{Lapidus.AJP.50.45,Julienne.Faraday.142.361} and the full Casimir-Polder potential~\cite{Nabutovski.JETP.50.352} as well as incoming boundary conditions for the description of inelastic collisions at the surface of the nanotube~\cite{Arnecke.PRA.75.042903,Friedrich.PRA.65.032902,Fink.PRA.81.062714}. 
The Casimir-Polder potential depends on the nontrivial dielectric properties $\varepsilon(\mathrm{i}\omega)$ of a multiwall CNT~\cite{Klimchitskaya.RevModPhys.81.1827,Klimchitskaya.JPhysA.39.6481,Blagov.PRB.71.235401}. Nonetheless, its long-range asymptote, which dominates the scattering process, depends solely on the static dielectric constant $\varepsilon(0)$.
The experimental data~\cite{Schneeweiss.NNano.2012.93} shown in fig.~\ref{fig:bec} (full circles) is perfectly reproduced by our calculations with a frequency-independent dielectric constant $\varepsilon$ in the range from $\varepsilon=2.5$ (dot-dashed line) up to $\varepsilon=100$ (dashed line), which seems to include the range of realistic values.\\

% Thermal Cloud of Atoms 
\section{Absorption of a thermal cloud}
\begin{figure}
\onefigure[width=75mm]{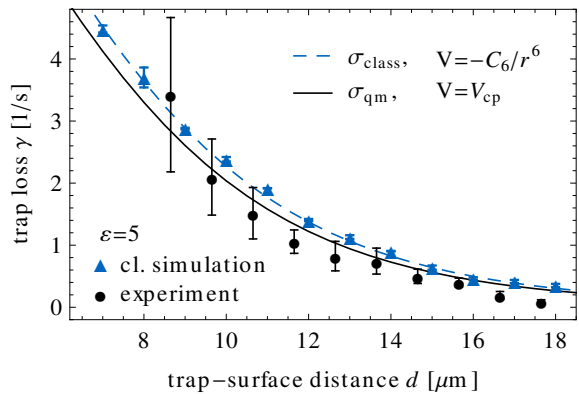}
\caption{(Color online) Trap loss $\gamma$ for a thermal cloud of rubidium atoms overlapping with a carbon nanotube. The black solid line shows the predicted trap loss for a dielectric constant of $\varepsilon=5$ which is in good agreement with the experimental results~\cite{Schneeweiss.NNano.2012.93} (full circles and errorbars). 
The blue triangles show a three-dimensional classical simulation of the system which reproduces the predicted classical trap loss (blue dashed line) and shows rather small deviations from the quantum mechanical results.}
\label{fig:tc}
\end{figure}

The trap loss for a thermal cloud of rubidium atoms is shown in fig.~\ref{fig:tc}. The velocity of the atoms in a thermal cloud is given by a Maxwell-Boltzmann distribution with $T=100$nK and the loss rate constant is given by
\begin{equation}\label{eq:abs_cs tc}
 K_\mathrm{in}^\mathrm{2D} = \left< v_{\bot}\sigma_\mathrm{abs}(v_{\bot}) \right>\,,
\end{equation}
where $v_{\bot}$ is the velocity in the plane perpendicular to the nanotube. The density of the atoms in the thermal cloud is given by a gaussian distribution with $\sigma_x=27\mu$m and $\sigma_y=\sigma_z=5.5\mu$m.

\begin{figure}
\onefigure[width=75mm]{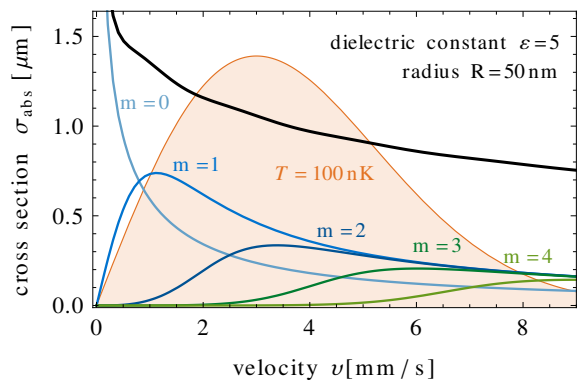}
\caption{(Color online) Absorption cross section for rubidium atoms scattered at a cylinder of radius $50$nm and a dielectric constant of $\varepsilon=5$. The black curve shows the full absorption cross section $\sigma_\mathrm{abs}$ including all contributing partial waves. The blue and green curves show the individual contribution of each partial wave from $m=0$ ($s$-wave) up to $m=4$ ($g$-wave). The orange region in the background shows the velocity distribution perpendicular to the tube (in arbitrary units) for a cloud of $100$nK. Notice that due to the reduced dimensionality of the system $\sigma_\mathrm{abs}$ is a length.}
\label{fig:abs_cs}
\end{figure}

The absorption cross section $\sigma_\mathrm{abs}$ in two dimensions for rubidium atoms scattered by an infinite cylinder of radius $50$nm and dielectric constant $\varepsilon=5$ is illustrated in fig.~\ref{fig:abs_cs} and has been calculated by well established scattering theory~\cite{Lapidus.AJP.50.45,Julienne.Faraday.142.361} using the full interaction potential~\cite{Nabutovski.JETP.50.352} and incoming boundary conditions~\cite{Arnecke.PRA.75.042903,Friedrich.PRA.65.032902,Fink.PRA.81.062714}. For a cloud with $T=100$nK, angular momenta up to the $g$-wave ($m=4$) contribute significantly to the absorption cross section $\sigma_\mathrm{abs}$ at a CNT with a radius of $50$nm, which is in contrast to simpler models~\cite{Schneeweiss.NNano.2012.93}.

Our prediction shown in fig.~\ref{fig:tc} (solid line), which is based on the full Casimir-Polder potential and a dielectric constant $\varepsilon=5$, is in good agreement with the experimental results (full circles). Variations of $\varepsilon$ in a wide range from $\varepsilon=2.5$ up to $\varepsilon=100$ lead to rather small variations of our results compared to the experimental error bars (see fig.~\ref{fig:tc_ret}).

For the case of a thermal cloud we also performed a time dependent three-dimensional classical simulation of a cloud of $1000$ atoms at constant temperature in a harmonic trap, overlapping with an absorbing nanotube. The interaction potential between the atoms and the tube is approximated by the long-range asymptotics of the Casimir-Polder potential. The effect of the finite length of the tube has been taken into account by a smooth decrease of the potential at the endings of the tube. The loss rate observed in this simulation (triangles in fig.~\ref{fig:tc}) is reproduced by the result predicted via eqs.~(\ref{eq:gamma total}) and (\ref{eq:abs_cs tc}) using the corresponding classical absorption cross section for the asymptotic potential (dashed line in fig.~\ref{fig:tc}). The comparison of these results with the full quantum mechanical calculation and the experimental data shows a good agreement. Thus, in contrast to the extremely nonclassical BEC [see eq.~(\ref{eq:abs_cs bec})], the behaviour of the 
thermal cloud at $T=100$nK is essentially determined by classical dynamics.\\

% Role of Retardation
\section{Influence of retardation}
\begin{figure}
\onefigure[width=75mm]{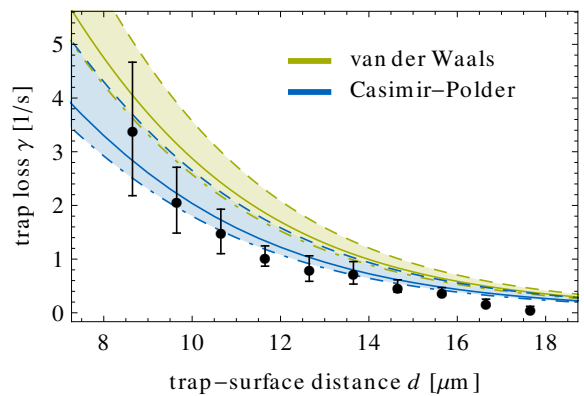}
\caption{(Color online) Trap loss $\gamma$ for a thermal cloud of rubidium atoms. The loss rates has been calculated via eq.~(\ref{eq:gamma total}) based on the full Casimir-Polder potential (lower curves) and on the van der Waals potential only, where retardation effects are neglected (upper curves). The shaded area shows loss rates for dielectric constant of the nanotube in a regime from $\varepsilon=2.5$ (dot-dashed line) up to $\varepsilon=100$ (dashed line). The solid line shows the trap loss for a dielectric constant of $\varepsilon=5$. The experimental data~\cite{Schneeweiss.NNano.2012.93} (full circles) clearly deviates from the prediction calculated with the van der Waals potential only.}
\label{fig:tc_ret}
\end{figure}

A fundamental question that needs to be addressed in future experiments is the influence of retardation on scattering processes for such systems. It has recently been shown that for realistic atom-nanotube systems, retardation effects are important in the case of a perfectly conducting tube~\cite{Fink.PRA.85.040702}.
This result holds for the present case of a carbon-nanotube immersed in a cloud of thermal atoms; the nonretarded van der Waals potential fails to describe the trap loss (upper curves in fig.~\ref{fig:tc_ret}), within a wide range of dielectric constants from $\varepsilon=2.5$ (dot-dashed line) up to $\varepsilon=100$ (dashed line). The results obtained with the full Casimir-Polder potential (lower curves in fig.~\ref{fig:tc_ret}) match the experimental data within the same range of dielectric constants.
For a BEC the differences of the loss rates obtained from the van der Waals potential and the full Casimir-Polder potential are rather small and not resolvable in this range of dielectric properties.\\

% Summary
\section{Conclusion}
In this Letter, an accurate parameter-free ab-initio description of the trap loss in a hybrid system consisting of a single nanotube immersed in an ultracold atomic quantum gas, such as a thermal cloud or a BEC is presented. 
A quantum mechanical calculation of the absorption cross section based on the exact Casimir-Polder potential leads to perfect agreement with recent experiments~\cite{Schneeweiss.NNano.2012.93}, both for a thermal cloud and for a BEC. Furthermore, it has been shown that the van der Waals potential fails to reproduce the loss rates of a thermal cloud of atoms; retardation effects that are accounted for in the Casimir-Polder potential need to be considered.
An accurate description of this hybrid system is achieved with a proper use of scattering theory together with the exact Casimir-Polder potential. The present approach gives insight into the underlying processes and therefore promotes a deeper understanding of hybrid systems which is essential for the design of future nanodevices.\\

% Acknowledgements
\acknowledgments
The authors thank J.~Fort\'agh, A.~G\"unther, P.~Schneeweiss, T.~E.~Judd for providing experimental parameters and for helpful discussions. Useful input by H.~Friedrich is gratefully acknowledged. This work was supported by the Deutsche Forschungsgemeinschaft, Az.:~ FR591/13-2.

\end{document}